\documentclass{article}
\usepackage{spconfa4,amsmath,graphicx}
\usepackage{multicol,multirow,booktabs}
\usepackage[citecolor=black,
             colorlinks=true,
             linkcolor=black,
             urlcolor  = black,
             pdftitle={},
             pdfauthor={},
             pdfkeywords={},
             pdfsubject={}]{hyperref}
\usepackage{xcolor}
\usepackage{acronym}
\usepackage{amssymb}

\usepackage[english]{babel}
\addto\extrasenglish{}
\addto\extrasenglish{}
\addto\extrasenglish{}
\addto\extrasenglish{}
\addto\extrasenglish{}
\addto\extrasenglish{}

\title{Challenges of Multi-Speaker Extraction for Real Conversational Speech Enhancement}
\name{Robert Sutherland$^1$\thanks{This work was supported by the UKRI AI Centre for Doctoral Training in Speech and Language Technologies (SLT) and their Applications funded by UK Research and Innovation [grant number EP/S023062/1]. For the purpose of open access, the author has applied a Creative Commons Attribution (CC BY) licence to any Author Accepted Manuscript version arising. This work was also supported by WS Audiology.}, Stefan Goetze$^{1,2}$, Jon Barker$^1$}
\address{$^1$School of Computer Science, University of Sheffield, Sheffield, United Kingdom\\
            $^2$South Westphalia University of Applied Sciences, Iserlohn, Germany}

\acrodef{AHD}       {assistive hearing device}
\acrodef{CT}        {close-talk}
\acrodef{ECHI}      {Enhancing Conversations to address Hearing Impairment}
\acrodef{FiLM}      {feature-wise linear modulation}
\acrodef{fwSegSNR}  {frequency-weighted segmental SNR}
\acrodef{HA}        {hearing aid}
\acrodef{MSX}       {multi-speaker extraction}
\acrodef{NN}        {neural network}
\acrodef{PESQ}      {perceptual evaluation of speech quality}
\acrodef{VAD}       {voice-activity detection}
\acrodef{SNR}       {signal-to-noise ratio}
\acrodef{STFT}      {short-time Fourier transform}
\acrodef{STOI}      {short-time objective intelligibility}
\acrodef{TSX}       {target-speaker extraction}
\begin{document}
\ninept
\maketitle

\begin{abstract}
    Target-speaker and multi-speaker extraction are techniques for extracting speech from a desired speaker or desired speakers in the presence of other speakers and/or noise. Neural network approaches for this task are often trained and evaluated using simulated datasets, with balanced amounts of target speech and speaker enrolment samples which closely match the target speech. However, in real multi-party conversations, participants are often silent for more time than they are speaking, and their enrolment speech samples can differ substantially from the target speech in the conversation. These factors can impact the training and evaluation of these techniques on recordings of real conversations. This work proposes a new loss function, which helps mitigate the effect of excess silence in training, improving STOI from 0.55 to 0.60, and frequency-weighted segmental SNR from 4.35 to 5.12. Additionally, the impact of the mismatch between the enrolment speech and target speech is explored.
\end{abstract}

\section{Introduction}

While existing \acp{AHD}, such as \acp{HA}, can improve the listening experience of hearing-impaired listeners in many scenarios, their benefit is often reduced in complex acoustic environments, such as cafes and restaurants. In particular, those with hearing impairments can find it hard to socialise in multi-party conversations in these challenging environments.

The tasks of \ac{TSX} have \ac{MSX} have been developed to tackle exactly this scenario with \ac{NN}-based techniques \cite{cornell2023multi,hao2024if, serre2025mtse}. However, these techniques often operate under assumptions which are rarely valid in the context of \ac{AHD} processing. This work highlights and addresses some of these concerns.

While \ac{TSX} techniques have successfully been applied to low-latency \ac{HA} processing~\cite{cornell2023multi}, they only focus on a single speaker, while conversations may include a number of speakers; while this technique could be applied iteratively (once per target speaker), this is inefficient, especially in the context of low-latency processing.

This motivates the task of \ac{MSX}, where models extract more than one conversation partner at any given time~\cite{serre2025mtse, ao2024used}. While this is more aligned with the multi-party conversation scenario, existing techniques do not consider the strict latency requirements of \ac{AHD} systems; those listening to audio through a \ac{HA} find delays of $>20$ ms disturbing~\cite{stone1999tolerable}, so any \ac{NN} designed for \acp{AHD} must satisfy this constraint.

Additionally, \ac{MSX} techniques are often explored using simulated datasets, with controlled speaker activity and overlaps~\cite{wichern2019wham, cosentino2020librimix}. In multi-party conversations, speaker activity is very varied, with challenging distributions of utterance lengths and overlaps~\cite{cetin2006analysis}; for example, in a four-party conversation, an interlocutor could be silent for as much as $70\%$ of the conversation, creating an imbalance between speech and silence which can impact the training of \ac{NN} models~\cite{zhang2023speaker}.

A further mismatch occurs with the target speaker information; \ac{TSX}/\ac{MSX} techniques typically use enrolment speech to identify the target speaker(s) e.g. by computing speaker embeddings~\cite{cornell2023multi,ao2024used}. With simulated data, this enrolment speech closely matches the target speech in the noisy mixture. In real conversations, the enrolment speech may be read speech, which has different acoustic properties to the spontaneous speech in conversations~\cite{summers1988effects, howell1991comparison}.

This work uses the CHiME-9 ECHI dataset~\cite{sutherland2025echi}, which captures real four-party conversations in a simulated noisy environment, to investigate the effect of training \ac{TSX}/\ac{MSX} models in a realistic scenario. With this, a new loss function is proposed to mitigate the effect of the imbalance in speaker activity by randomly applying \ac{VAD}-masking to the loss function, controlling the amount of silence that the model is exposed to, leading to improvements in objective speech intelligibility and speech quality metrics.

Further, it is observed that system performance degrades when there is a greater mismatch between the speaker embedding of the enrolment sample and the speaker embedding obtained from the target speaker in conversation.

\section{Methodology}

This section first describes the data used for this work in \autoref{ssec:Data}, followed by an introduction to the signal model for the \ac{TSX} and \ac{MSX} tasks in \autoref{ssec:SignalModel}. Then, a technical description of the \ac{NN} structure is given in \autoref{ssec:NetworkStructure}, followed by a mathematical description of the newly proposed loss function in \autoref{ssec:LossFunction}.

\subsection{Data}
\label{ssec:Data}

\begin{figure}[!ht]
    \centering
    \includegraphics[width=0.8\linewidth]{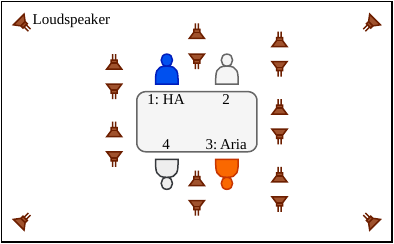}
    \caption{The approximate layout of the ECHI recording scenario (not drawn to scale). This example layout shows the blue participant in position $1$ wearing the \acp{HA} and the orange participant in position $3$ wearing the Aria glasses.}
    \label{fig:schematic}
\end{figure}

This work uses the ECHI dataset~\cite{sutherland2025echi}, which comprises recordings of real four-party conversations in a simulated noisy environment. In the recording sessions, one person (out of four conversation partners) was wearing two $2$-channel \ac{HA} shells (one on each ear), and another person was wearing Project Aria smart glasses~\cite{engel2023project}, which record $7$-channel audio. A schematic of the recording setup is given in \autoref{fig:schematic}.

Each participant additionally wore a \ac{CT} microphone to capture speech close to the mouth of the wearer. Reference signals provided in the dataset for the \acp{HA} and Aria glasses were constructed by denoising the \ac{CT} microphones~\cite{defossez2020real} and applying a delay to account for sound propagation~\cite{sutherland2025echi}. Also included in the dataset are clean recordings of each participant reading the first paragraph from the rainbow passage~\cite{fairbanks1960}, which can be used as enrolment speech samples of the target speakers to help extract them from the noisy mixture.

The dataset provides $48$ conversation recording sessions, each with a duration of $\sim 36$ minutes; these are split into \emph{train} ($30$ sessions), \emph{development} ($10$ sessions) and \emph{evaluation} ($8$ sessions). Participants in each set are disjoint, so speakers in the train set only appear in the train set, and similarly for development and evaluation.

\subsection{Signal Model}
\label{ssec:SignalModel}

The noisy audio recordings at the microphone positions of either the \ac{HA} shells or Project Aria smart glasses of different participants in the conversation are denoted by
\begin{equation}
    \mathbf{X} = \mathbf{S}_1 + \mathbf{S}_2 + \mathbf{S}_3 + \mathbf{S}_4 + \mathbf{N},
\end{equation}
where $\mathbf{S}_i\in\mathbb{R}^{C\times T}$ is the multi-channel recording of the speech of speaker $i$ and $\mathbf{N}\in\mathbb{R}^{C\times T}$ is the multi-channel noise, as picked up by the $C$-channel recording device for a duration of $T$ samples. For the \ac{HA} microphones $C=4$, and for the Aria glasses $C=7$. In the following, it is assumed for simplicity of notation that the wearer is participant $1$, and the targets are participants $2$, $3$ and $4$ (cf.~\autoref{fig:schematic}).

For training and objective evaluation, the reference signals $\mathbf{T}=\left[\mathbf{t}_2,\mathbf{t}_3,\mathbf{t}_4\right]\in\mathbb{R}^{3\times T}$ denote the single-channel reference speech for participants $2$, $3$ and $4$ (participant $1$ is ignored, as this is the device wearer). The rainbow passage recordings, which are used as the enrolment utterances, are denoted $\mathbf{R}=\left\{\mathbf{r}_2,\mathbf{r}_3,\mathbf{r}_4\right\}\in\mathbb{R}^{3\times K}$.

Finally, the speech enhancement \acp{NN}, $\mathcal{M}$, are defined for the \ac{TSX} and \ac{MSX} tasks as
\begin{align}
    \mathcal{M}_\text{TSX}\left(\mathbf{X},\mathbf{r}_i\right)&=\mathbf{\hat{t}}_i,\\
    \mathcal{M}_\text{MSX}\left(\mathbf{X},\mathbf{R}\right) &= \mathbf{\hat{T}},
\end{align}
so $\mathcal{M}_\text{TSX}$ aims to extract one target speaker at a time and $\mathcal{M}_\text{MSX}$ aims to extract all three conversation partners at once.

\subsection{Network Architecture}
\label{ssec:NetworkStructure}

\begin{figure}[!ht]
    \centering
    \includegraphics[width=0.9\linewidth]{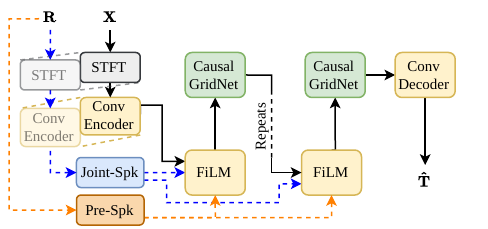}
    \caption{System architecture for $\mathcal{M}_\mathrm{MSX}$. The enrolment speech $\mathbf{R}$ can be processed by a jointly-trained speaker encoder (Joint-Spk) or a pre-trained speaker encoder (Pre-Spk).}
    \label{fig:model}
\end{figure}

In keeping with typical \ac{TSX}/\ac{MSX} systems~\cite{cornell2023multi,hao2024if,serre2025mtse}, the model comprises two main components: a speaker encoder network, which produces speaker embeddings of the target speaker(s), and a speech enhancement network, which produces the audio of the target speaker(s). Specifically, this model uses a speech extraction TF-GridNet model, as this has been effective for \ac{TSX} with low-latency processing~\cite{cornell2023multi,hao2024if}. The architecture diagram is shown for $\mathcal{M}_\text{MSX}$ in \autoref{fig:model}; for $\mathcal{M}_\text{TSX}$, one enrolment recording, $\mathbf{r}_i$, is be provided, and a single channel of audio $\mathbf{\hat{t}}_i$ produced.

\subsubsection{Input Features}

First, $\mathbf{X}$ is downsampled to $16$ kHz before the complex spectrogram is computed. A real-valued representation of the complex spectrogram is obtained by concatenating the real and imaginary components along the channel dimension. For the jointly-trained speaker encoder, the spectrogram of $\mathbf{R}$ is computed in the same way, but the pre-trained speaker encoder operates on the raw waveform~\cite{jung2022pushing}. For the \ac{TSX} case, the single-channel spectrogram of $\mathbf{r}_i$ is concatenated in a new dimension.

The \ac{STFT} uses a Hanning window, with a window size of $128$ samples ($8$ ms), and the hop size is $64$ samples ($4$ ms).

\subsubsection{Speaker Encoder}

The purpose of the speaker encoder is to produce an embedding of the target speaker, based on some enrolment speech samples; here, the rainbow passage provided in the ECHI dataset~\cite{sutherland2025echi} is used. This work considers two approaches: joint training of a speaker embedding model~\cite{cornell2023multi,hao2024if}, or using a pre-trained model~\cite{jung2022pushing, jung2024espnet}. It should be noted that while \ac{AHD} algorithms are required to be low-latency, the outputs of this speaker embedding model can be cached for inference, meaning they need not be low-latency.

For the jointly-trained model, the network architecture follows \cite{hao2024if}. The spectrogram of the enrolment speech is passed through the same convolutional encoder as the noisy audio, before being processed by a sequence of U-Nets~\cite{ronneberger2015unet}. Finally, a $2$D average pooling is applied along the time-dimension to form a speaker embedding $\mathbf{v}\in\mathbb{R}^L$.

The pre-trained model takes the same clean-speech samples, but processes them through the RawNet3 speaker embedding model~\cite{jung2022pushing, jung2024espnet}, to produce a $1$D speaker embedding.

For \ac{MSX}, the embeddings for each of the enrolment speech recordings are computed separately, and concatenated when conditioning the \ac{FiLM} layer~\cite{perez2017visual}.


\subsubsection{Speech Enhancement Network}
\begin{figure}[!ht]
    \centering
    \includegraphics[width=1\linewidth]{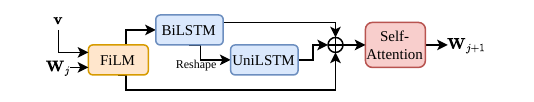}
    \caption{The GridNet architecture with FiLM conditioning.}
    \label{fig:filmgrid}
\end{figure}

The backbone of the speech enhancement system is the TF-GridNet architecture~\cite{wang2023tf,cornell2023multi}, along with a \ac{FiLM} layer~\cite{perez2017visual}. First, the multi-channel spectrogram passes through a convolutional encoder to produce the representation $\mathbf{W}_0$. The \ac{FiLM} layer conditions $\mathbf{W}_0$ on the speaker embedding $\mathbf{v}$, before a Bi\-LSTM operates along the feature dimension. The output is then reshaped, so the unidirectional LSTM operates over the time dimension, before the input $\mathbf{W}_0$ and both LSTM outputs are summed and passed to the self-attention block, which outputs $\mathbf{W}_1$. This block can be repeated, hence \autoref{fig:filmgrid} takes $\mathbf{W}_j$ as input and produces $\mathbf{W}_{j+1}$.

This block must be low-latency, so specific choices were made to ensure minimal look-ahead: the time-dimension LSTM is unidirectional, and the self-attention mechanism is masked to avoid any look-ahead to future frames.

\subsection{Loss Function}
\label{ssec:LossFunction}

The loss function proposed here is based on a linear combination of spectral distances~\cite{arik2018fast}, which uses a \emph{spectral convergence} term
\begin{equation}
    \label{eqn:specconv}
    \mathcal{L}_\text{SC}\left(\mathbf{\hat{T}},\mathbf{T}\right)=\frac{\parallel\mid\text{STFT}(\mathbf{T})\mid - \mid\text{STFT}(\mathbf{\hat{T}})\mid\parallel_F}{\parallel\mid\text{STFT}(\mathbf{T})\mid\parallel_F},
\end{equation}
and a \emph{magnitude spectrogram distance}
\begin{equation}
    \label{eqn:magspecdist}
    \begin{split}
        \mathcal{L}_\text{Mag}\left(\mathbf{\hat{T}},\mathbf{T}\right)=\parallel&\log(\mid\text{STFT}(\mathbf{T})\mid+\epsilon) \\&-\log(\mid\text{STFT}(\mathbf{\hat{T}})\mid+\epsilon)\parallel_1,
    \end{split}
\end{equation}
with $\parallel\cdot\parallel_F$ denoting the Frobenius norm, $\parallel\cdot\parallel_1$ the L$1$-norm, \linebreak $\mid\cdot\mid$ the absolute value of a complex number, and $\epsilon=1\times10^{-8}$ is a small value to avoid $\log(0)$. Equations (\ref{eqn:specconv}) and (\ref{eqn:magspecdist}) are written for the \ac{MSX} case, but can equally be computed for the \ac{TSX} case using $\mathbf{\hat{t}}_i$ and $\mathbf{t}_i$.

A spectral loss was chosen over a time-domain loss, such as \ac{SNR}, as the latter are highly sensitive to the lack of precise sample alignment of the reference signal. When computing the \ac{STFT} of the audio for the loss function, the best-performing window size was found to be $1024$ samples, corresponding to $64$ ms at a sampling frequency of $16$ kHz, which is ample to absorb any of the sample-level misalignment.

The final spectral loss is a linear combination of (\ref{eqn:specconv}) and (\ref{eqn:magspecdist}):

\begin{equation}
    \label{eqn:finaloss}
    \mathcal{L}_\text{Spec}\left(\mathbf{\hat{T}},\mathbf{T}\right) = \mathcal{L}_\text{SC}(\mathbf{\hat{T}},\mathbf{T}) + \mathcal{L}_\text{Mag}(\mathbf{\hat{T}},\mathbf{T}).
\end{equation}

To compensate for the high proportion of silence in each target speaker's channel, the loss function can be masked using \ac{VAD}, so that only the loss during speech segments is computed, as described by (\ref{eqn:vad}).

\begin{align}
    \text{VAD}\left(\mathbf{T}, \mathbf{V}\right) &= \mathbf{T} \odot \mathbf{V},\\
    \mathcal{L}_\text{VAD}\left(\mathbf{\hat{T}},\mathbf{T},\mathbf{V}\right) &= \mathcal{L}_\text{Spec}\left(\text{VAD}(\mathbf{\hat{T}}, \mathbf{V}),\text{VAD}(\mathbf{T},\mathbf{V})\right)
    \label{eqn:vad}
\end{align}
where $\mathbf{V}\in\{0,1\}^{N\times T}$ is the \ac{VAD} mask, with $0$ indicating silence and $1$ indicating speech activity for each of the $N$ target speakers, and $\odot$ denotes the Hadamard product. This formulation ensures that the loss is computed only for time segments where the respective target speaker is active.

The loss function $\mathcal{L}_\text{VAD}$ simply ignores any portion of the signal where the target in each channel is silent. However, the intention here is to control the amount of silence computed in the loss, not to ignore it completely. This leads to the randomly masked loss function

\begin{equation}
    \label{eqn:vadloss}
    \mathcal{L}_\text{VAD}^{(x)}(\mathbf{\hat{T}},\mathbf{T}, \mathbf{V}) =
    \begin{cases}
        \mathcal{L}_\text{VAD}(\mathbf{\hat{T}},\mathbf{T},\mathbf{V}),x\%\text{ of batches} \\
        \mathcal{L}_\text{Spec}(\mathbf{\hat{T}}, \mathbf{T}),\quad \ (100-x)\%\text{ of batches},
    \end{cases}
\end{equation}
where for each batch, a random choice is made that ensures each loss $\mathcal{L}_\text{VAD}$ and $\mathcal{L}_\text{Spec}$ are used on the specified proportion of batches. When applying this \ac{VAD}-masked loss function, careful consideration must be made about how often the loss function is masked; this is discussed in further detail in \autoref{ssec:vadscores}.

\section{Training Setup}

All training and model hyperparameters were consistent across the model variants, with the exception of the learning rates. The final speech enhancement network had $3.8$M trainable parameters. The jointly-trained speaker encoder network had $4.3$M trainable parameters, while the pre-trained RawNet3 speaker embedding model had $28.5$M frozen parameters. The \ac{VAD} masks $\mathbf{V}$ are generated using speech timestamps provided in the ECHI dataset~\cite{sutherland2025echi}.

Models were trained for $30$ epochs using the Adam optimiser~\cite{kingma2014adam}. The learning rate was scheduled using a linear warm-up over the first three epochs, before cosine annealing was applied over the remaining $27$ epochs~\cite{loshchilov2016sgdr}, and gradients were clipped to $1$. The best learning rate for each model variant was either $1\times10^{-4}$ or $5\times10^{-5}$.

A validation loop on the development set was run every two epochs, and the checkpoint used for the final evaluation was the one which produced the highest \ac{STOI}~\cite{taal2010stoi} score on the development set. Further details of the exact training configurations can be found in the GitHub repository\footnote{Found at: \href{https://github.com/MxRobsta/ECHI-SpeechEnhancement}{github.com/MxRobsta/ECHI-SpeechEnhancement}}.

Using a single NVIDIA H100 GPU, training typically took $\sim1.5$ days and enhancement for the $4.8$ hours of audio in the evaluation set took $\sim 45$ minutes.

\section{Results}
\label{sec:Results}

To evaluate the performance of the system, a range of speech metrics is used. These metrics are only computed on speech segments of the audio.


To assess speech intelligibility, \ac{STOI}~\cite{taal2010stoi} is computed. The speech quality metrics used are \ac{PESQ}~\cite{rix2001pesq} and the composite measures~\cite{hu2008speechmet}, which compute \emph{Csig} (signal quality), \emph{Cbak} (background intrusion), and \emph{Covl} (overall quality for communication). Finally, \ac{fwSegSNR}~\cite{tribolet1978speech} is computed as a signal level metric; this metric is derived from \ac{SNR} with a weighting applied to the frequency bands to more closely reflect human hearing.

The results are presented in \autoref{tab:results} and are computed for the $6525$ target speech segments for the ECHI \ac{HA} recordings. Similar trends are observed on the ECHI Aria glasses recordings; these results are available online at the link above.

\begin{table}[!ht]
    \caption{Results on the evaluation split of the ECHI dataset. The Pre-Spk column indicates whether the speaker embedding is pre-trained (\checkmark) or jointly trained (blank).}
    \resizebox{1\linewidth}{!}{
        \begin{tabular}{ccccccccc}
            \toprule
            Model      & Pre-Spk        & Loss                                  & fwSegSNR          & STOI              & PESQ              & Csig              & Cbak              & Covl \\
            \midrule
            \multicolumn{3}{c}{Noisy Audio}                                     & 0.37              & 0.55              & 1.10              & 1.51              & 1.07              & 1.20\\
            \multicolumn{3}{c}{CHiME-9 ECHI Baseline}                           & 3.39              & 0.53              & 1.10              & 1.69              & 1.08              & 1.30\\
            \midrule
            TSX        &                &$\mathcal{L}_\text{VAD}^{(80)}$        & 3.55              & 0.54              & 1.14              & 1.82              & 1.11              & 1.38\\
            TSX        & \checkmark     &$\mathcal{L}_\text{VAD}^{(80)}$        & 3.88              & 0.56              & 1.15              & 1.88              & 1.11              & 1.42\\
            \midrule
            MSX        &                &$\mathcal{L}_\text{Spec}$              & 4.05              & 0.55              & 1.11              & 1.51              & 1.08              & 1.22\\
            MSX        & \checkmark     &$\mathcal{L}_\text{Spec}$              & 4.35              & 0.54              & 1.15              & 1.75              & 1.18              & 1.36\\
            \midrule
            MSX        &                & $\mathcal{L}_\text{VAD}^{(80)}$       & 4.13              & \textbf{0.60}     & 1.18              & 1.92              & 1.44              & 1.44\\
            MSX        & \checkmark     & $\mathcal{L}_\text{VAD}^{(80)}$       & \textbf{5.12}     & \textbf{0.60}     & \textbf{1.19}     & \textbf{1.97}     & \textbf{1.55}     & \textbf{1.48}\\
            \bottomrule
        \end{tabular}
    }
    \label{tab:results}
\end{table}

For $\mathcal{M}_\text{TSX}$, training with $\mathcal{L}_\text{Spec}$ was found to annihilate the signal, producing silence. This is due to the high proportion of silence in the reference signal, so this model was only trained with $\mathcal{L}^{(80)}_\text{VAD}$. This effect was avoided for $\mathcal{M}_\text{MSX}$ as there was typically speech present in at least one channel, meaning the overall target was rarely just silence.

The first point to note is that across all the metrics, the best \ac{TSX} model performs similarly to the best \ac{MSX} model trained with $\mathcal{L}_\text{Spec}$ from (\ref{eqn:finaloss}). This is interesting, as the \ac{MSX} model is provided with more information about the overall scene; it receives speaker information about all of the target speakers, which should make the task easier.

However, it is clear that training the \ac{MSX} models with $\mathcal{L}_\text{VAD}^{(80)}$ improves the performance across all of the metrics ($p<0.01$). Further, training with the pre-trained speaker embeddings shows a slight improvement over all metrics except \ac{STOI} ($p<0.01$ except for \ac{STOI}, $p>0.1$); significance is determined by Wilcoxon test. This suggests that high proportions of silence in the reference audio can limit performance, but randomly applying the \ac{VAD}-mask can compensate for this and improve performance.

Further details on tuning the proportion of silence presented to the model are discussed in \autoref{ssec:vadscores}, and the influence of the pre-trained speaker embeddings is considered in \autoref{ssec:spksim}.

\subsection{VAD-Masked Loss Function}
\label{ssec:vadscores}

The argument behind defining the randomly \ac{VAD}-masked loss in (\ref{eqn:vadloss}) is that the amount of silence that the model trains with can be controlled. For this to be meaningful, the threshold used to mix the masked loss with the raw loss must be carefully considered. To assess this, \autoref{tab:vadmask} shows the same metrics as in \autoref{tab:results}, computed with different degrees of \ac{VAD}-masking in (\ref{eqn:vadloss}). Note that the models presented here are $\mathcal{M}_\text{MSX}$ with pre-trained speaker embeddings, and the scores for $0\%$ masking are equivalent to training with $\mathcal{L}_\text{Spec}$ from  (\ref{eqn:finaloss}), and $100\%$ is equivalent to training only with $\mathcal{L}_\text{VAD}$ from (\ref{eqn:vad}). 

\begin{table}[!ht]
    \caption{Scores for different levels of usage of $\mathcal{L}_\text{VAD}$. * indicates scores which appeared in \autoref{tab:results}.}
    \centering
    \begin{scriptsize}
        \begin{tabular}{ccccccc}
            \toprule
            VAD     & fwSegSNR & STOI & PESQ & Csig & Cbak & Covl \\
            \midrule
            0\%*    & 4.35          & 0.54          & 1.15          & 1.75          & 1.18          & 1.36 \\
            20\%    & 4.20          & 0.55          & 1.15          & 1.67          & 1.28          & 1.31 \\
            40\%    & 4.76          & 0.58          & 1.18          & 1.91          & 1.55          & 1.45 \\
            60\%    & 4.53          & 0.59          & 1.17          & 1.84          & 1.39          & 1.40 \\
            80\%*   & \textbf{5.12} & 0.60          & 1.19          & 1.97          & 1.55          & 1.48 \\
            100\%   & 5.02          & \textbf{0.62} & \textbf{1.21} & \textbf{2.06} & \textbf{1.56} & \textbf{1.54} \\
            \bottomrule
        \end{tabular}
    \end{scriptsize}
    \label{tab:vadmask}

\end{table}
The results in \autoref{tab:vadmask} show that training with too much silence in the reference signals degrades the \ac{MSX} model's performance on objective metrics. For all metrics, the worst scores occur with either $0\%$ or $20\%$ \ac{VAD}-masking, and the best scores with $80\%$ or $100\%$ \ac{VAD}-masking.

There are small performance increases for the $100\%$ model over the $80\%$ model in the perceptual metrics. This can be expected to a certain extent; these metrics are only computed on speech segments, so training only on speech segments is likely to produce better results here. However, as described previously, it is also important that the model can process scenarios where one or more of the target speakers are not speaking. This suggests that while training on exclusively using $\mathcal{L}_\text{VAD}$ may yield slight improvements for these objective metrics, for the overall use-case, $80\%$ would provide a better experience. 

\subsection{Speaker Similarity}
\label{ssec:spksim}

The speaker embeddings are used to help guide the \ac{TSX} and \ac{MSX} systems towards the target speakers. These embeddings are generated using clean, read speech samples of the target speakers, but it is known that there are acoustic differences between read speech and the spontaneous speech that can be found in the conversations~\cite{summers1988effects,howell1991comparison}.

To assess the impact of this mismatch, RawNet3 embeddings~\cite{jung2022pushing} are computed from two speech sources. The enrolment embedding is generated using the enrolment speech, $\mathbf{r}_i$ from the ECHI dataset~\cite{sutherland2025echi}, as is used during training. A reference embedding is computed by taking the mean of five embeddings computed from the reference signal, $\mathbf{t}_i$; these were generated by randomly selecting five speech segments with a duration of more than $3$ seconds. The difference between embeddings is computed with cosine similarity.

To expand the number of data points, speakers in both the development and evaluation sets are considered, giving $n=54$ speakers. By computing Pearson-$r$ correlations, it can be seen that \ac{PESQ} shows a medium correlation with the cosine similarity ($r=0.49$, $p<0.01$), and all other metrics show a strong correlation ($r>0.5$, $p<0.01$).

This suggests that the mismatch in target speaker information is causing a drop in performance. Some speakers are poorly matched from their enrolment speech to the target speech, and in these cases, the objective performance is worse. 

\section{Conclusion}

The real-time enhancement of conversations provides slightly different challenges when compared to more traditional \ac{TSX} and \ac{MSX} tasks. This work shows that using a \ac{VAD}-masked loss function leads to significant improvements in both perceptual and signal-based metrics, suggesting that models trained with this methodology will perform better. It can also be seen that using pre-trained speaker embeddings can provide some marginal benefit for these metrics.

Future work could consider using a speaker adaptivity mechanism to update speaker embeddings during a conversation, meaning that the speakers can be more easily recognised by their spontaneous speech instead of their read speech.

\bibliographystyle{IEEEtran}
\bibliography{refs}

\end{document}